\documentclass[preprint]{revtex4-2}
\usepackage[utf8]{inputenc}
\usepackage[T1]{fontenc}

\usepackage[greek, brazil, english]{babel}
\usepackage{amsmath, amssymb, bm}
\usepackage{amsfonts} 
\usepackage{yhmath}
\usepackage{bbold}   
\usepackage{slashed}
\usepackage{mathrsfs} 
\usepackage{lmodern}
\usepackage{mathtools} 
\usepackage{semantic} 

\usepackage[a4paper, margin=1.5cm]{geometry}
\usepackage{makeidx} 
\usepackage{indentfirst} 

\usepackage{graphicx, subfigure} 
\usepackage{tabularray}

\usepackage{hyperref}
\hypersetup{
    colorlinks = true,
    linkcolor  = blue,
    citecolor  = blue,           
    filecolor  = magenta,
    urlcolor   = cyan,
}

\usepackage{printlen} 
\usepackage{calculator} 
\usepackage{ifthen} 

\usepackage{tikz} 
\usetikzlibrary{arrows.meta} 
\usepackage{siunitx} 
\usepackage{setspace} 
\makeatletter
\AtBeginDocument{
    \check@mathfonts
    \fontdimen13\textfont2=4.5pt 
    \fontdimen14\textfont2=4.5pt 
    \fontdimen15\textfont2=4.2pt 
    \fontdimen16\textfont2=2.8pt 
    \fontdimen17\textfont2=2.8pt 
}
\makeatother

\usepackage{color, transparent}	

\newcommand{\AdS}{\mathrm{AdS}}

\newcommand{\hor}{\mathrm{hor}}
\newcommand{\horp}{\mathrm{hor},\,p}
\newcommand{\onshell}{\text{on shell}}

\renewcommand{\Re}{\textrm{Re}}
\renewcommand{\Im}{\textrm{Im}}

\newcommand{\bvec}[1]{\ensuremath{\bm{#1}}}

\let\mgt\gg 
\let\mlt\gg 

\renewcommand{\gg}{\bvec{g}}

\newcommand  {\xx}{\bvec{x}}

\renewcommand{\cal}[1]{\ensuremath{\mathcal{#1}}}

\newcommand{\calE}{\cal{E}}

\newcommand{\calL}{\cal{L}}

\let\originalleft\left
\let\originalright\right
\renewcommand{\left}{\mathopen{}\mathclose\bgroup\originalleft}
\renewcommand{\right}{\aftergroup\egroup\originalright}

\DeclareSymbolFont{mylargesymbols}{OMX}{ccex}{m}{n}
\SetSymbolFont{mylargesymbols}{bold}{OMX}{cmex}{m}{n}
\DeclareMathDelimiter{\lbrace}{\mathopen}{symbols}{"66}{mylargesymbols}{"08}
\DeclareMathDelimiter{\rbrace}{\mathclose}{symbols}{"67}{mylargesymbols}{"09}
\DeclareMathDelimiter{(}{\mathopen}{operators}{"28}{mylargesymbols}{"00}
\DeclareMathDelimiter{)}{\mathclose}{operators}{"29}{mylargesymbols}{"01}
\DeclareMathDelimiter{[}{\mathopen}{operators}{"5B}{mylargesymbols}{"02}
\DeclareMathDelimiter{]}{\mathclose}{operators}{"5D}{mylargesymbols}{"03}
\DeclareMathSymbol{\braceld}{\mathord}{mylargesymbols}{"7A}
\DeclareMathSymbol{\bracerd}{\mathord}{mylargesymbols}{"7B}
\DeclareMathSymbol{\bracelu}{\mathord}{mylargesymbols}{"7C}
\DeclareMathSymbol{\braceru}{\mathord}{mylargesymbols}{"7D}

\newcommand{\PAR}[2][-1]{
    \MULTIPLY{0.75}{#1}{\vara}
    \ADD{\vara}{0.50}{\varb}
    \ifthenelse
    { \equal{#1}{-1} }
    { \ensuremath{ {\left( #2 \right) } } }
    { \ensuremath{
            {               \left( \rule{0pt}{ \varb ex} \right. \hspace{-1pt} }
            #2
            { \hspace{-1pt} \left. \rule{0pt}{ \varb ex} \right)               }
} } }

\newcommand{\COL}[2][-1]{
    \MULTIPLY{0.75}{#1}{\tempa}
    \ADD{\tempa}{0.50}{\tempb}
    \ifthenelse
    { \equal{#1}{-1} }
    { \ensuremath{ {\left[\, #2 \,\right] } } }
    { \ensuremath{
            {               \left[ \rule{0pt}{ \tempb ex } \right. \hspace{-1pt} }
            \,#2\,
            { \hspace{-1pt} \left. \rule{0pt}{ \tempb ex } \right]               }
} } }

\newcommand{\LCOL}[1][-1]{%
    \MULTIPLY{0.75}{#1}{\tempa}
    \ADD{\tempa}{0.50}{\tempb}
    \ensuremath{  \hspace{-1pt} \left[ \rule{0pt}{ \tempb ex} \right. \hspace{-1pt} }
}

\newcommand{\RCOL}[1][-1]{%
    \MULTIPLY{0.75}{#1}{\tempa}
    \ADD{\tempa}{0.50}{\tempb}
    \ensuremath{ \hspace{-1pt} \left. \rule{0pt}{ \tempb ex} \right] \hspace{-1pt} }
}

\newcommand{\CHA}[2][-1]{
    \MULTIPLY{0.75}{#1}{\tempa}
    \ADD{\tempa}{0.50}{\tempb}
    \ifthenelse
    { \equal{#1}{-1} }
    { \ensuremath{ {\left\{ #2 \right\} } } }
    { \ensuremath{
            {              \left\{ \rule{0pt}{ \tempb ex} \right.  \hspace{-1pt} }
            #2
            { \hspace{-1pt} \left. \rule{0pt}{ \tempb ex} \right\}               }
} } }

\newcommand{\LCHA}[1][-1]{%
    \MULTIPLY{0.75}{#1}{\tempa}
    \ADD{\tempa}{0.50}{\tempb}
    \ensuremath{ \hspace{-1pt} \left\{ \rule{0pt}{ \tempb ex} \right.  \hspace{-1pt} }
}

\newcommand{\RCHA}[1][-1]{%
    \MULTIPLY{0.75}{#1}{\tempa}
    \ADD{\tempa}{0.50}{\tempb}
    \ensuremath{ \hspace{-1pt} \left.  \rule{0pt}{ \tempb ex} \right\} \hspace{-1pt} }
}

\newcommand{\at}[2][-1]{
    \MULTIPLY{0.75}{#1}{\tempa}
    \ADD{\tempa}{0.50}{\tempb}
    \ifthenelse
    { \equal{#1}{-1} }
    { \ensuremath{ {\left.\, #2 \,\right| } } }
    { \ensuremath{
            {           \!\!\left. \rule{0pt}{ \tempb ex} \right. \hspace{-1pt} }
            \,#2\,
            { \hspace{-1pt} \left. \rule{0pt}{ \tempb ex} \right|              }
} } }

\newcommand{\abs}[2][-1]{
    \MULTIPLY{0.75}{#1}{\tempa}
    \ADD{\tempa}{0.50}{\tempb}
    \ifthenelse
    { \equal{#1}{-1} }
    { \ensuremath{ {\left| #2 \right| } } }
    { \ensuremath{
            {               \left| \rule{0pt}{ \tempb ex} \right. \hspace{-1pt} }
            #2
            { \hspace{-1pt} \left. \rule{0pt}{ \tempb ex} \right|               }
} } }

\newcommand{\bra}[2][0]{
    \MULTIPLY{0.75}{#1}{\tempa}
    \ADD{\tempa}{0.50}{\tempb}
    \ifthenelse
    { \equal{#1}{-1} }
    { \ensuremath{ {\left< #2 \right| } } }
    { \ensuremath{
        {               \left< \rule{0pt}{ \tempb ex} \right. \hspace{-1pt} }
        #2
        { \hspace{-1pt} \left. \rule{0pt}{ \tempb ex} \right|               }
} } }

\newcommand{\ket}[2][0]{
    \MULTIPLY{0.75}{#1}{\tempa}
    \ADD{\tempa}{0.50}{\tempb}
    \ifthenelse
    { \equal{#1}{-1} }
    { \ensuremath{ {\left| #2 \right> } } }
    { \ensuremath{
        {               \left| \rule{0pt}{ \tempb ex} \right. \hspace{-1pt} }
        #2
        { \hspace{-1pt} \left. \rule{0pt}{ \tempb ex} \right>               }
} } }

\newcommand{\braket}[3][0]{
    \MULTIPLY{0.75}{#1}{\tempa}
    \ADD{\tempa}{0.50}{\tempb}
    \ifthenelse
    { \equal{#1}{-1} }
    { \ensuremath{ {\left< #2 | #3 \right> } } }
    { \ensuremath{
        {               \left< \rule{0pt}{ \tempb ex} \right. \hspace{-1pt} }
        #2
        {               \left| \rule{0pt}{ \tempb ex} \right. \hspace{-1pt} }
        #3
        { \hspace{-1pt} \left. \rule{0pt}{ \tempb ex} \right>               }
} } }

\newcommand{\ketbra}[3][0]{
    \MULTIPLY{0.75}{#1}{\tempa}
    \ADD{\tempa}{0.50}{\tempb}
    \ifthenelse
    { \equal{#1}{-1} }
    { \ensuremath{         \ket{#2}\!\bra{#3}         } }
    { \ensuremath{ \ket[\tempb]{#2}\!\bra[\tempb]{#3} } }
}

\newcommand{\average}[2][0]{
    \MULTIPLY{0.75}{#1}{\tempa}
    \ADD{\tempa}{0.50}{\tempb}
    \ifthenelse
    { \equal{#1}{-1} }
    { \ensuremath{ {\left< #2 \right> } } }
    { \ensuremath{
        {               \left< \rule{0pt}{ \tempb ex} \right. \hspace{-1pt} }
        #2
        { \hspace{-1pt} \left. \rule{0pt}{ \tempb ex} \right>               }
} } }

\newcommand{\ddfrac}[2]{\dfrac{\raisebox{-1pt}{$\displaystyle #1$}}{\displaystyle {#2}^{\vphantom{\rule{0pt}{5pt}}}} }

\newcommand{\complexi}{\hspace{.15ex}\textrm{i}\hspace{.15ex}}
\newcommand{\exponentiale}{\hspace{.15ex}\textrm{e}\hspace{.15ex}}

\makeatletter

\@namedef{u8:\detokenize{α}}{\alpha}
\@namedef{u8:\detokenize{β}}{\beta}
\@namedef{u8:\detokenize{γ}}{\gamma}
\@namedef{u8:\detokenize{δ}}{\delta}
\@namedef{u8:\detokenize{ϵ}}{\epsilon}
\@namedef{u8:\detokenize{ε}}{\varepsilon}
\@namedef{u8:\detokenize{ζ}}{\zeta}
\@namedef{u8:\detokenize{η}}{\eta}
\@namedef{u8:\detokenize{θ}}{\theta}
\@namedef{u8:\detokenize{ϑ}}{\vartheta}
\@namedef{u8:\detokenize{κ}}{\kappa}
\@namedef{u8:\detokenize{λ}}{\lambda}
\@namedef{u8:\detokenize{μ}}{\mu}
\@namedef{u8:\detokenize{ν}}{\nu}
\@namedef{u8:\detokenize{ξ}}{\xi}
\@namedef{u8:\detokenize{π}}{\pi}
\@namedef{u8:\detokenize{ρ}}{\rho}
\@namedef{u8:\detokenize{ϱ}}{\varrho}
\@namedef{u8:\detokenize{σ}}{\sigma}
\@namedef{u8:\detokenize{ς}}{\varsigma}
\@namedef{u8:\detokenize{τ}}{\tau}
\@namedef{u8:\detokenize{υ}}{\upsilon}
\@namedef{u8:\detokenize{ϕ}}{\phi}
\@namedef{u8:\detokenize{φ}}{\varphi}
\@namedef{u8:\detokenize{χ}}{\chi}
\@namedef{u8:\detokenize{ψ}}{\psi}
\@namedef{u8:\detokenize{ω}}{\omega}
\@namedef{u8:\detokenize{ϖ}}{\varpi}

\@namedef{u8:\detokenize{Γ}}{\Gamma}
\@namedef{u8:\detokenize{Δ}}{\Delta}
\@namedef{u8:\detokenize{Θ}}{\Theta}
\@namedef{u8:\detokenize{Λ}}{\Lambda}
\@namedef{u8:\detokenize{Ξ}}{\Xi}
\@namedef{u8:\detokenize{Π}}{\Pi}
\@namedef{u8:\detokenize{Σ}}{\Sigma}
\@namedef{u8:\detokenize{ϒ}}{\Upsilon}
\@namedef{u8:\detokenize{Φ}}{\Phi}
\@namedef{u8:\detokenize{Ψ}}{\Psi}
\@namedef{u8:\detokenize{Ω}}{\Omega}

\@namedef{u8:\detokenize{±}}{\pm}
\@namedef{u8:\detokenize{²}}{^{2}}
\@namedef{u8:\detokenize{³}}{^{3}}
\@namedef{u8:\detokenize{¹}}{^{1}}
\@namedef{u8:\detokenize{⁰}}{^{0}}
\@namedef{u8:\detokenize{·}}{\hspace{-0.05ex}\cdot\hspace{-0.05ex}}
\@namedef{u8:\detokenize{×}}{\hspace{-0.05ex}\times\hspace{-0.05ex}}
\@namedef{u8:\detokenize{î}}{\complexi}
\@namedef{u8:\detokenize{ĩ}}{\complexi}
\@namedef{u8:\detokenize{ħ}}{\hbar}
\@namedef{u8:\detokenize{ḏ}}{\mathrm{d}}
\@namedef{u8:\detokenize{ē}}{\exponentiale}
\@namedef{u8:\detokenize{†}}{\dagger}
\@namedef{u8:\detokenize{‡}}{\ddagger}
\@namedef{u8:\detokenize{ı}}{\imath}
\@namedef{u8:\detokenize{ℓ}}{\ell}
\@namedef{u8:\detokenize{←}}{\leftarrow}
\@namedef{u8:\detokenize{→}}{\rightarrow}
\@namedef{u8:\detokenize{∂}}{\partial}
\@namedef{u8:\detokenize{≠}}{\neq}
\@namedef{u8:\detokenize{≤}}{\leq}
\@namedef{u8:\detokenize{≥}}{\geq}
\@namedef{u8:\detokenize{∞}}{\infty}
\@namedef{u8:\detokenize{∫}}{\int}
\@namedef{u8:\detokenize{≈}}{\approx}
\@namedef{u8:\detokenize{≡}}{\equiv}
\@namedef{u8:\detokenize{□}}{\Box}
\@namedef{u8:\detokenize{ⵈ}}{\cdots}

\makeatother

\expandafter\newcommand\csname u8:\detokenize{√}\endcsname[2][]{%
    \ifthenelse
    { \equal{#1}{} }
    { \sqrt{#2} }
    { \sqrt[#1]{#2} }
}

\expandafter\newcommand\csname u8:\detokenize{ˍ}\endcsname[1][]{%
    \ifthenelse
    { \equal{#1}{} }
    { \phantom{0} }
    { \hspace{#1} }
}

\expandafter\newcommand\csname u8:\detokenize{˷}\endcsname[1]{%
    \phantom{#1}
}

\mathlig{<<}{\mlt}
\mathlig{>>}{\mgt}
\mathlig{!=}{\neq}
\mathlig{=!}{\neq}
\mathlig{/=}{\neq}
\mathlig{=/}{\neq}
\mathlig{<>}{\neq}
\mathlig{><}{\neq}
\mathlig{==}{\equiv}
\mathlig{='}{\simeq}
\mathlig{<=}{\leq}
\mathlig{=<}{\leq}
\mathlig{>=}{\geq}
\mathlig{=>}{\geq}

\mathlig{+-}{\pm}
\mathlig{-+}{\mp}

\mathlig{-->}{\rightarrow}
\mathlig{--->}{\longrightarrow}
\mathlig{<--}{\leftarrow}
\mathlig{<---}{\longleftarrow}
\mathlig{<->}{\leftrightarrow}
\mathlig{<-->}{\longleftrightarrow}
\mathlig{|->}{\mapsto}
\mathlig{|-->}{\longmapsto}
\mathlig{==>}{\Rightarrow}
\mathlig{===>}{\Longrightarrow}
\mathlig{==>}{\Leftarrow}
\mathlig{===>}{\Longleftarrow}
\mathlig{<=>}{\leftrightarrow}
\mathlig{<==>}{\longleftrightarrow}

\begin{document}

\preprint{}

\title{Bottomonium  Dissociation in a Rotating Plasma}

\author{Nelson R. F. Braga}
\affiliation{Instituto de Física, Universidade Federal do Rio de Janeiro,   RJ 21941-909, Brazil.}

\author{Yan F. Ferreira}
\affiliation{Instituto de Física, Universidade Federal do Rio de Janeiro,   RJ 21941-909, Brazil.}

\date{\today}

\begin{abstract}
Heavy vector mesons provide important information about the quark gluon plasma (QGP)  formed in heavy ion collisions. This happens because the fraction of quarkonium states that are produced depends on the properties of the medium. The intensity of the dissociation process in a plasma is affected by  the temperature, the chemical potential   and  the presence of magnetic fields. These effects have been studied by many authors in the recent years.  Another important factor that can affect the dissociation of heavy mesons, and still lacks of a better understanding, is the rotation of the plasma. Non central collisions form a plasma with angular momentum. Here we use a holographic model to investigate the thermal spectrum of bottomonium quasi-states in a rotating medium in order to describe how a non vanishing angular velocity affects the dissociation process. 
\end{abstract}


\maketitle

\newpage

\vspace{.75\baselineskip}
\section{Introduction}

Heavy ion collisions, produced in particle accelerators, lead to the formation of a new state of matter, a plasma where quarks and gluons are deconfined. This so called QGP behaves like a perfect fluid and lives for a very short time\cite{Bass:1998vz, Scherer:1999qq, Shuryak:2008eq, Casalderrey-Solana:2011dxg}.
The study of this peculiar state of matter is based on the analysis of the particles that are observed after the hadronization process occur and the plasma disappears. For this to be possible, it is necessary to understand how the properties of the QGP, like temperature ($T$)  and density ($\mu$), affect the spectra that reach the detectors. 
In particular, quarkonium states, like bottomonium,  are very interesting since they survive the deconfinement process that occur when the QGP is formed. They undergo a partial dissociation, with an intensity that depends on the characteristics of the medium, like $T$ and $ \mu$. So, it is important to find it out how the properties of the plasma affect the dissociation. 

Bottomonium quasi-states in a thermal medium can be described using  holographic models \cite{Braga:2015jca,Braga:2016wkm, Braga:2017oqw,Braga:2017bml,Braga:2018zlu, Braga:2018hjt,Braga:2019yeh} see also \cite{MartinContreras:2021bis, Zollner:2021stb,Mamani:2022qnf}. In particular, the improved holographic model proposed in \cite{Braga:2017bml}, which will be considered here, involves three energy parameters: one representing the heavy quark mass, another associated with the intensity of the strong interaction (string tension) and another with the non-hadronic decay of quarkonium.  This model provides good estimates for masses and decay constants.  

Besides temperature, density and magnetic fields, 
another, less studied, property that affects the thermal behaviour is the rotation of the QGP, that occurs in non-central collisions. For previous works about the rotation effects in the QGP, see for example \cite{Miranda:2014vaa,Jiang:2016wvv,McInnes:2016dwk,Mamani:2018qzl,Wang:2018sur,Chernodub:2020qah,Arefeva:2020jvo,Chen:2020ath,Zhou:2021sdy,Braguta:2021jgn,Braguta:2021ucr,Golubtsova:2021agl,Fujimoto:2021xix,Braga:2022yfe,Chen:2022mhf,Golubtsova:2022ldm,Zhao:2022uxc,Chernodub:2022}. In particular, a holographic description of the QGP in rotation can be found in Refs.      \cite{Chen:2020ath,Braga:2022yfe}. 
A rotating plasma with uniform rotational speed is described holographically in these works by a rotating black hole with cylindrical symmetry. Rotation is obtained by a coordinate transformation and the holographic model obtained predicts that plasma rotation decreases the critical temperature of confinement/deconfinement transition \cite{Braga:2022yfe}. For a very recent study of charmonium in a rotating plasma see \cite{Zhao:2023pne}. 

The purpose of this work is to study how rotation of the plasma affects the thermal spectrum of bottomonium quasi-states. In other words, we want to understand what is the effect of rotation in the dissociation process of $ b \bar b $. We will follow two complementary approaches. One is to calculate the thermal spectral functions and the other is to find the quasinormal models associated with bottomonium in rotation.

The organization is the following. In section II we present a holographic model for bottomonium in a rotating plasma.  In III we work out the equations of motion for the fields that describe the quasi states. In IV we discuss the solutions, taking into account the incoming wave boundary conditions on the black hole horizon. In section V we calculate the spectral functions for bottomonium and in VI we present the complex frequencies of the quasi-normal modes. Finally, section VII contains our conclusions and discussions about the results obtained.

\vspace{.75\baselineskip}
\section{Holographic Model For Quarkonium in the Plasma}
\label{sec: model}

Vector mesons are represented holographically by a vector field $V_m  = (V_t, V_1, V_2, V_3, V_z)$ , that lives, in the case of a non rotating plasma,  in a five dimensional anti-de Sitter ($\AdS_5$) black hole space with metric
\begin{gather}
    ḏs² = \ddfrac{R²}{z²}
          \PAR[3]{\!-f(z) ḏt² + (ḏx^1)² + (ḏx^2)² + (ḏx^3)² + \ddfrac{1}{f(z)} ḏz²},
    \label{eq: non-rotating metric with cartesian coordinates}
\shortintertext{with}
    f(z) = 1 - \ddfrac{z^4}{z_h^4}\,.
\end{gather}
 The constant $R$ is the $\AdS$ radius and the Hawking temperature of the black hole, given by
\begin{gather}
    T = \ddfrac{1}{π z_h},
    \label{eq: temperature without rotation}
\end{gather}
is identified with the temperature of the plasma.

The  action integral for the field has the form
\begin{gather}
    I = ∫ḏ^4x ∫_{0}^{z_h} ḏz\, √{-g}\, \calL,
    \label{eq: action}
\shortintertext{with the Lagrangian density}
    \calL = -\dfrac{1}{4 g_5^2} ē^{-ϕ(z)} g^{mp} g^{nq} F_{mn} F_{pq},
\end{gather}
where $F_{mn} = \nabla_{\!m} V_n - \nabla_{\!n} V_m = ∂_m V_n - ∂_n V_m$, with $\nabla_{\!m}$ being the covariant derivative.  The dilaton like background 
field $ϕ(z)$ is \cite{Braga:2017oqw,Braga:2017bml}
\begin{gather}
    ϕ(z) = κ^2 z^2 + Mz + \tanh\!\PAR[3]{\dfrac{1}{Mz} - \dfrac{κ}{√{Γ}}}.
\end{gather}
The back reaction of $ϕ(z)$ on the metric is not taken into account. This modified dilaton is introduced in order to obtain an  approximation for the values of masses and decay constants of heavy vector mesons.   The three parameters of the model are fixed at zero temperature to give the best values for these quantities, specially the decay constants, compared with the experimental values for bottomonium found in the particle data group table \cite{ParticleDataGroup:2020ssz}:  $κ_b = \SI{2.45}{\giga\eV}$, $√{\smash[b]{Γ_b}} = \SI{1.55}{\giga\eV}$ and $M_b = \SI{6.2}{\giga\eV}$. They give the results presented in table \ref{tab: bottomonium masses and decay constants}.

\begin{table}[htb]
\centering
\begin{tblr}{
    colspec = {m{1.5cm}m{3.1cm}m{3.1cm}m{3.1cm}m{3.1cm}},
    columns = {c},
    stretch = 0,
    rowsep = 4pt,
    cell{1}{1} = {c=5}{c},
    hlines = {0.5pt},
    vlines = {0.5pt},
}
    Bottomonium Masses and Decay Constants \\
    State &
    Experimental Masses ($\si{\mega\electronvolt}$) &
    Masses on the tangent model ($\si{\mega\electronvolt}$) &
    Experimental Decay Constants ($\si{\mega\electronvolt}$) &
    Decay Constants on the tangent model ($\si{\mega\electronvolt}$)
    \\
    $1S$  & $ ˍ9460.30 ± 0.26 $ & $ˍ6905$  & $ 715.0  ± ˍ4.8  $ & 719 \\
    $2S$  & $ 10023.26 ± 0.31 $ & $ˍ8871$  & $ 497.4  ± ˍ4.5  $ & 521 \\
    $3S$  & $ 10355.2ˍ ± 0.5ˍ $ & $10442$  & $ 430.1  ± ˍ3.9  $ & 427 \\
    $4S$  & $ 10579.4ˍ ± 1.2ˍ $ & $11772$  & $ 341ˍ\; ± 18ˍ\; $ & 375
\end{tblr}\\[10pt]
\caption{Comparison of bottomonium masses and decay constants obtained experimentally \cite{ParticleDataGroup:2020ssz} and from the tangent model.}
\label{tab: bottomonium masses and decay constants}
\end{table}

In order to have some characterization of the quality of the fit, one can define the root mean square percentage error (RMSPE) as
\begin{gather}
    \textrm{RMSPE}
    = 100\% × √{ \ddfrac{1}{N-N_p} \sum_{i\,=\,1}^{N}
                                           \PAR[3]{\ddfrac{y_i-\hat{y}_i}{\hat{y}_i}}^{\!\!2} },
\end{gather}
where $N = 8$ is the number of experimental points (4 masses and 4 decay constants), $N_p = 3$ is the number of parameters of the model, the $y_i$'s are the values of masses and decay constants predicted by the model and the $\hat{y}_i$'s are the experimental values of masses and decay constants. With this definition, we have $\textrm{RMSPE} = 14.8\%$ for bottomonium.

Now, in order to analyse the case of a rotating plasma, with homogeneous angular velocity, we consider  an $AdS_5$ space with cylindrical symmetry by writing the metric as
\begin{gather}
    ḏs² = \ddfrac{R²}{z²}
          \PAR[3]{\!-f(z) ḏt² + ℓ²ḏφ² + (ḏx^1)² + (ḏx^2)² + \ddfrac{1}{f(z)} ḏz²},
    \label{eq: non-rotating metric}
\end{gather}
where $R$ is again the AdS radius and $ℓ$ is the hypercylinder radius.

As in ref. \cite{Zhou:2021sdy,Chen:2020ath,Braga:2022yfe}, we introduce rotation via the Lorentz-like coordinate transformation
\begin{gather}
\begin{aligned}
    t &--> γ (t + Ω ℓ² φ) \\
    φ &--> γ (Ω t + φ),
\end{aligned}
    \label{eq: coordinate transformation}
\shortintertext{with}
    γ = \ddfrac{1}{√{1-Ω²ℓ²}},
\end{gather}
where $Ω$ is the angular velocity of the rotation.
With this transformation, the metric \eqref{eq: non-rotating metric} becomes
\begin{gather}
\begin{split}
    ḏs^2
    = \ddfrac{R^2}{z^2}
      \LCOL[3] -   γ² \PAR[2]{ f(z) - Ω²ℓ²  } ḏt^2 
               + 2 γ² \PAR[2]{   1 - f(z)   } Ωℓ² ḏtḏφ
               +   γ² \PAR[2]{ 1 - Ω²ℓ²f(z) } ℓ² ḏφ^2 \qquad\\
               + \; (ḏx^1)^2 + (ḏx^2)^2
               + \ddfrac{1}{f(z)} ḏz^2
      \RCOL[3],
\end{split}
\label{eq: metric}
\end{gather}
The temperature of the rotating AdS black hole is\cite{Zhou:2021sdy,Chen:2020ath,Braga:2022yfe}  
\begin{gather}
    T = \dfrac{1}{π z_h} √{1 - Ω²ℓ²}.
    \label{eq: temperature}
\end{gather}
Note that we recover \eqref{eq: non-rotating metric} by doing $Ω --> 0$ in \eqref{eq: metric}.

We assume that the rotating black hole metric of Eq. \eqref{eq: metric} is dual to a cylindrical slice of rotating plasma is dual to a cylindrical slice of rotating plasma.  This interpretation can be justified by analysing the angular momentum $J$. For the metric \eqref{eq: metric}, $J$ can be calculated   \cite{Braga:2022yfe} with the result
 
\begin{equation}
J = -\frac{\partial \Phi}{\partial \Omega}=\frac{2L^3 }{\kappa^2}\frac{\Omega}{z_{h}^{4}(1-\Omega^2l^2)}\;, 
\end{equation}
 while for the metric \eqref{eq: non-rotating metric} one has $ J = 0$. Thus, the coordinate transformation is adding angular momentum to the system and therefore representing a plasma in rotation. 



\vspace{.75\baselineskip}
\section{Equations of Motion}
\label{sec: Equations of Motion}

By extremizing the action \eqref{eq: action}, we find the equations of motion
\begin{gather}
    ∂_n (√{-g} ē^{-ϕ} F^{mn}) = 0.
    \label{eq: eqs of motion in compact form}
\end{gather}

We now choose a Fourier component of the field and, for simplicity, consider zero momentum (meson at rest), $V_m(t, \xx, z) = v_m(ω,z) ē^{-îωt}$. We also choose the gauge $V_z = 0$. The equations of motion \eqref{eq: eqs of motion in compact form} become
\begin{gather}
    \ddfrac{ 1 - Ω²ℓ²f }{ 1 - Ω²ℓ² } \ddfrac{ω²}{f²} v_i
    + \PAR[3]{ \ddfrac{f'}{f} - \ddfrac{1}{z} - ϕ' } v_i'
    + v_i''
    = 0 \qquad (i = 1, 2),
        \label{eq: eq motion m=1,2}
    \\[0.3\baselineskip]
    \begin{multlined}[b][.8\textwidth]
    - \COL[3]{  \ddfrac{1}{1 - Ω²ℓ²f^{-1}} \ddfrac{f'}{f}
              + \ddfrac{1 - f}{f - Ω²ℓ²} \PAR{\ddfrac{1}{z} + ϕ'} } Ωℓ² v_t'
    + \ddfrac{1-f}{f-Ω²ℓ²} Ω ℓ² v_t'' \\
    + \ddfrac{ 1 - Ω²ℓ² }{ 1 - Ω²ℓ²f^{-1} } \ddfrac{ω²}{f²} v_φ
    + \COL[3]{ \ddfrac{1}{1 - Ω²ℓ²f^{-1}}\ddfrac{f'}{f} - \ddfrac{1}{z} - ϕ' } v_φ'
    + v_φ''
    = 0,
    \end{multlined}
        \label{eq: eq motion m=φ}
    \\[0.3\baselineskip]
    \begin{multlined}[b][.8\textwidth]
    \phantom{-}\,\,
    \COL[3]{  \ddfrac{1}{f^{-1} - Ω²ℓ²} \ddfrac{f'}{f}
            + \ddfrac{1 - f}{1 - Ω²ℓ²f} \PAR{\ddfrac{1}{z} + ϕ'} } Ω v_φ'
    - \ddfrac{1-f}{1-Ω²ℓ²f} Ω v_φ'' \\
    - \COL[3]{ \ddfrac{1}{(Ω²ℓ²f)^{-1} - 1} \ddfrac{f'}{f} + \ddfrac{1}{z} + ϕ' } v_t'
    + v_t''
    = 0\;
    \end{multlined}
        \label{eq: eq motion m=t}
\shortintertext{and}
    v_t' - \ddfrac{1-f}{1-Ω²ℓ²f} Ω v_φ'
    = 0.
        \label{eq: eq motion m=z}
\end{gather}
where the prime stands for the derivative with respect to $z$, $f^{-1} = 1/f$, and, for simplicity, we omit the dependence on $z$ of $f$ and $ϕ$ and the dependence on $(ω,z)$ of the fields $v_μ$. These equations are not all independent, if we substitute \eqref{eq: eq motion m=z} into \eqref{eq: eq motion m=t} we obtain an identity, and substituting the same equation into \eqref{eq: eq motion m=φ}, we obtain an equation for $v_φ$ only. With this, the system of equations of motion simplifies to
\begin{gather}
    \ddfrac{ 1 - Ω²ℓ²f }{ 1 - Ω²ℓ² } \ddfrac{ω²}{f²} v_i
    + \PAR[3]{ \ddfrac{f'}{f} - \ddfrac{1}{z} - ϕ' } v_i'
    + v_i''
    = 0 \qquad (i = 1, 2),
        \label{eq: eq motion v_1,2}
    \\[0.1\baselineskip]
    \ddfrac{ 1 - Ω²ℓ²f }{ 1 - Ω²ℓ² } \ddfrac{ω²}{f²} v_φ
    + \PAR[3]{ \ddfrac{1}{1 - Ω²ℓ²f}\ddfrac{f'}{f} - \ddfrac{1}{z} - ϕ' } v_φ'
    + v_φ''
    = 0,
        \label{eq: eq motion v_φ}
    \\[0.1\baselineskip]
    v_t' - \ddfrac{1-f}{1-Ω²ℓ²f} Ω v_φ'
    = 0.
\end{gather}

Note that when $Ωℓ = 0$, equations \eqref{eq: eq motion v_1,2} and \eqref{eq: eq motion v_φ} are the same and the two states are degenerate. Rotation breaks this degeneracy.

From this point on, we will divide our analysis in two cases,  according to three possible polarizations. The first case is for polarizations in directions $x^1$ and $x^2$, for which $(v_m) = (0,v,0,0,0)$ or $(v_m) = (0,0,v,0,0)$. The second case is for the polarization in direction $φ$, for which $(v_m) = (0,0,0,v,0)$.


\vspace{.75\baselineskip}
\section{Solving the Equations of Motion}
\label{sec: Solving the Equations of Motion}

\subsection{Near the horizon behavior of the solution}

By approximating the function $f$ as the first term of its power series at $z = z_h$, we write
\begin{gather}
    f(z) =' f'(z_h)(z-z_h)
    \qquad
    \text{(for $z =' z_h$)}
\end{gather}
and we see that, in this region, equations \eqref{eq: eq motion v_1,2} and \eqref{eq: eq motion v_φ} both take the form
\begin{gather}
    \dfrac{γ^2 ω^2}{\displaystyle\rule{0pt}{11.5pt}f'(z_h)^2(z - z_h)^2}\,v_{\hor}^{}(z)
    + \dfrac{1}{z - z_h}\,v_{\hor}'(z) + \,v_{\hor}''(z) = 0,
    \label{eq: eqs of motion near the horizon}
\end{gather}
which, in terms of the temperature, can be written as
\begin{gather}
    \dfrac{ω^2}{\displaystyle\rule{0pt}{11.5pt}(4πT)^2}\,v_{\hor}^{}(z)
    + (z - z_h)\,v_{\hor}'(z) + (z - z_h)^2\,v_{\hor}''(z) = 0.
    \label{eq: eq of motion near the horizon}
\end{gather}
The two solutions of this equation are
\begin{gather}
    \PAR[3]{ 1 - \dfrac{z}{z_h} }^{\!\!-îω/4πT}
    \qquad\qquad
    \text{and}
    \qquad\qquad
    \PAR[3]{ 1 - \dfrac{z}{z_h} }^{\!\!+îω/4πT}.
\end{gather}
The solution with the minus sign in the exponent corresponds to an infalling wave at the horizon, while the other, with positive sign, corresponds to an outgoing wave. This becomes clear if one change to the Regge-Wheeler tortoise coordinate, as explained in refs. \cite{Mamani:2013ssa,Miranda:2009uw}.

The black hole allows only infalling waves at the horizon. Therefore, the field that solves the complete equations of motion has to obey the condition
\begin{gather}
    v(z) =' A \PAR[3]{1-\ddfrac{z}{z_h}}^{\!\!-îω/4πT}
    \qquad
    \text{(for $z =' z_h$)},
    \label{eq: vhor}
\end{gather}
where $A$ is a normalization constant. The norm of the field will have no importance for us, then we can set $A = 1$.

In order to solve the complete equations of motion \eqref{eq: eq motion v_1,2} and \eqref{eq: eq motion v_φ} numerically, we translate the infalling wave condition at the horizon in two boundary conditions to be imposed at a point  $z = z_0$ close to the horizon:
\begin{gather}
    v(z_0) = v_{\horp}(z_0)
    \qquad\text{and}\qquad
    v'(z_0) = v_{\horp}'(z_0),
\shortintertext{with}
    v_{\horp}(z)
    = \PAR[3]{ 1 - \dfrac{z}{z_h} }^{\!\!-îω/4πT}
      \sum_{n=0}^{p} a_n \PAR[3]{1 - \dfrac{z}{z_h}}^{\!\!n}. 
    \label{eq: vhorp}
\end{gather}
The function  $v_{\horp}(z)$ is just the infalling wave expression \eqref{eq: vhor} (with $A$ set to 1) times a polynomial correction of order $p$ introduced for purposes of the numerical calculations. The coefficient $a_0$ is, of course, 1 and the other coefficients $a_n$ are determined by imposing the equation of motion \eqref{eq: eq motion v_1,2} or \eqref{eq: eq motion v_φ} with $v = v_{\horp}$ to be valid up to the order $p$. The coefficients $a_n$ are, therefore, the same for polarization in directions $x^1$ and $x^2$ but are different from the ones for polarization in the direction $φ$. The point $z_0$ is a choosen value of $z$, close to $z_h$, where the approximation $v(z) =' v_{\horp}(z)$ is valid. Using equation \eqref{eq: vhor}, the leading term of the field, for $z$ close to $z_h$, can be written as:
\begin{gather}
    \cos\!\COL[3]{\ddfrac{ω}{4πT}\ln\!\PAR[3]{1-\ddfrac{z}{z_h}}}
    - î \sin\!\COL[3]{\ddfrac{ω}{4πT}\ln\!\PAR[3]{1-\ddfrac{z}{z_h}}}.
\end{gather}

In the region of values of $z$ close to the horizon, very small changes in $z$ produce significant changes in $v(z)$, a problem for the numerical calculation. For this reason, the point $z_0$ cannot be chosen too close to $z_h$. This is the reason why we introduce the polynomial perturbation from \eqref{eq: vhorp} in the infalling condition. In this work, the value $z_0 = 0.9 z_h$ with 20 coefficients $a_n$ was sufficient. 

\begin{figure}[htb!]
    \centering
    \vspace{\baselineskip}
    \includegraphics[scale=.75]{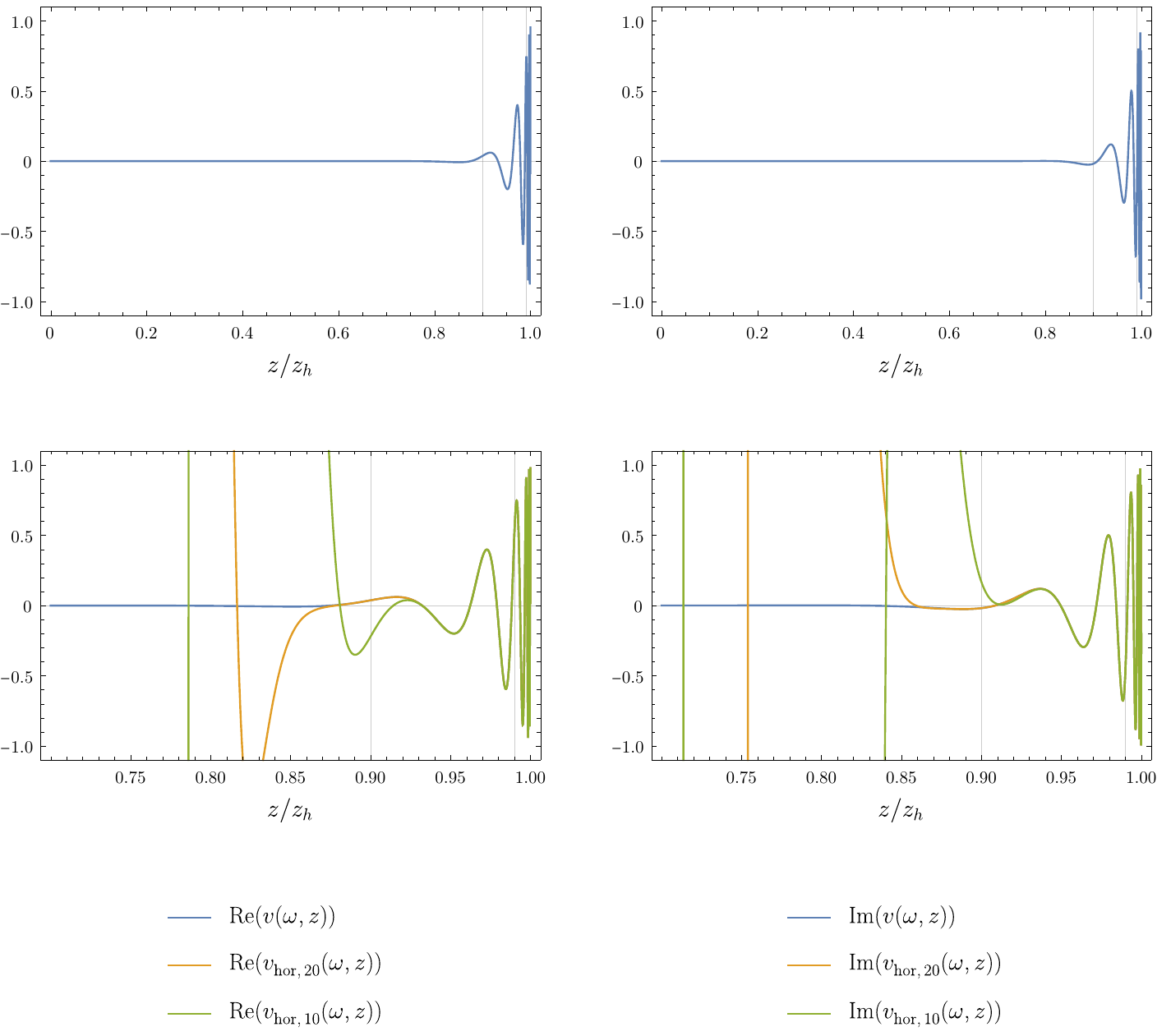}
    \caption{Real (left) and Imaginary (right) parts of the field $v(ω,z)$ (blue) and its approximations near the horizon $v_{\horp}(ω,z)$ with 20  (orange) and 10 (green) coefficients for the representative value of $ω = \SI{10}{\giga\electronvolt}$ for the non-rotating plasma at temperature $T = \SI{150}{\mega\electronvolt}$. In the first line we plot the field in all its domain. In the second line we zoom in the region from $z = 0.7 z_h$ to $z = z_h$. The vertical lines highlight the values $z = 0.9 z_h$ and $z = 0.99 z_h$.}
    \label{fig: field}
\end{figure}

Figure \ref{fig: field} shows the solution of the equation of motion for the non-rotating plasma at a specific temperature and for some representative value of $ω$ as well as two approximations for this solution near the horizon. From this figure, one can see that if we had chosen $z_0 = 0.99 z_h$, for example, instead of $z_0 = 0.9 z_h$, we would be in the unstable region and any small numerical error would be significantly propagated. Also, if we had used 10 coefficients, for example, instead of 20, we woldn't have a good aproximation for the field at $z_0 = 0.9 z_h$. For more discussion on this method, see, for example, \cite{Kaminski:2009ce}.

This method produces a numerical solution of equations \eqref{eq: eq motion v_1,2} or \eqref{eq: eq motion v_φ} with the infalling wave condition at $z_h$ for any value of $ω$. We will use these solutions to calculate spectral functions and quasinormal modes in the following sections.


\vspace{.75\baselineskip}
\section{Spectral Functions}
\label{sec: Spectral Functions}

Spectral Functions are defined in terms of the retarded Green's functions as
\begin{gather}
    ϱ_{μν}(ω) = -2\, \Im\,G_{μν}^R(ω).
    \label{def: spectral function}
\end{gather}
They provide an important way of analysing the dissociation of quarkonia in a thermal medium. At zero temperature, the spectral function of a quarkonium, considering just one particle states, is a set of delta peaks at the values of the holographic masses of table \ref{tab: bottomonium masses and decay constants}. At finite temperature, these peaks acquire a finite height and a non-zero width. As the temperature increases the height of each peak decreases and its width increases. This broadening effect of the peaks indicates dissociation in the medium. In this section we calculate the spectral function for bottomonium in a rotating plasma at three fixed temperatures in order to analyse the effect of the rotational speed in the dissociation process.

\subsection{Retarded Green's Function}

In the four dimensional vector gauge theory we define a retarded Green’s functions of the currents $J_ν$, that represent the heavy vector mesons, as
\begin{gather}
    G_{μν}^R = -î ∫ḏ^4x ē^{-îp·x} Θ(t) \average{[J_μ(x), J_ν(0)]}.
\end{gather}

The Son-Starinets prescription \cite{Son:2002sd} provide a way of extracting the retarded Green's function from the on shell action of the dual vector fields in AdS space
\begin{align}
    I_\onshell
    &= -\ddfrac{1}{4 g_5^2} ∫ḏ^4x∫_0^{z_h} ḏz\, √{-g}\, ē^{-ϕ(z)} F_{mn} F^{mn} \nonumber\\
    &= -\ddfrac{1}{2 g_5^2} ∫ḏ^4x∫_0^{z_h}ḏz\, ∂_m (√{-g}\, ē^{-ϕ(z)} V_n F^{mn}),
\end{align}
where we have used the equations of motion, \eqref{eq: eqs of motion in compact form}, to go from the first line to the second.

In the gauge $V_z = 0$, we have $V_n F^{mn} = V_ν F^{mν}$ and, therefore,
\begin{align}
    I_\onshell
    &= -\ddfrac{1}{2g_5^2}
    \begin{multlined}[t][.7\textwidth]
    \LCOL[3.5]
    \at[2.5]{∫ ḏx¹ḏx²ḏφḏz √{-g}\, ē^{-ϕ(z)} V_ν F^{tν}}_{t\,=\,-∞}^{t\,=\,+∞}
    + \at[2.5]{∫ ḏtḏx²ḏφḏz\, √{-g}\, ē^{-ϕ(z)} V_ν F^{1ν}}_{x^1\,=\,-∞}^{x^1\,=\,+∞} \\[0pt]
    + \at[2.5]{∫ ḏtḏx¹ḏφḏz\, √{-g}\, ē^{-ϕ(z)} V_ν F^{2ν}}_{x^2\,=\,-∞}^{x^2\,=\,+∞}\;
    + \at[2.5]{∫ ḏtḏx¹ḏx²ḏz\, √{-g}\, ē^{-ϕ(z)} V_ν F^{φν}}_{φ\,=\,0}^{φ\,=\,2π} \\[0pt]
    + \at[2.5]{∫ ḏ^4x\, √{-g}\, ē^{-ϕ(z)} V_ν F^{zν}}_{z\,=\,0}^{z\,=\,z_h}
    \,\RCOL[3.5].
    \end{multlined}
\end{align}
Since any physical field has to go to zero as $t$, $x^1$ or $x^2$ goes to $±∞$, the first three terms vanish. The point with $φ = 0$ is equivalent to the point $φ = 2π$, therefore, the fourth term vanishes too. This leaves us just with the surface term
\begin{align}
    I_\onshell
    &= -\ddfrac{1}{2g_5^2}
        \at[2.5]{∫ ḏ^4x\, √{-g}\, ē^{-ϕ(z)} V_ν F^{zν}}_{z\,=\,0}^{z\,=\,z_h}.
\end{align}
In momentum space and considering the meson at rest, we find
\begin{gather}
    I_\onshell
    = -\ddfrac{1}{2g_5^2} ∫ḏω\, \at[2.5]{ √{-g}\, ē^{-ϕ(z)}
                               g^{zz} g^{μν} v_μ(-ω,z) v_ν'(ω,z) }_{z\,=\,0}^{z\,=\,z_h}.
\end{gather}
Using the equation of motion \eqref{eq: eq motion m=z} to substitute $v_t'$ in terms of $v_φ'$, one eliminates $v_t$ ending up with
\begin{gather}
    \begin{multlined}[c][0.9\textwidth]
    I_\onshell
    = -\ddfrac{1}{2g_5^2}
       ∫ḏω\, √{-g}\, ē^{-ϕ(z)} g^{zz} \,
           \LCHA[3] \sum_{j=1,2} g^{jj} v_j(-ω,z) v_j'(ω,z) \\[-0.3\baselineskip]
                    + \COL[3]{\ddfrac{(g^{tφ})²}{-g^{tt}} + g^{φφ}} v_φ(-ω,z) v_φ'(ω,z) \RCHA[3]
           \at[4]{}_{z\,=\,0}^{z\,=\,z_h}.
    \end{multlined}
    \label{eq: on shell action without vt}
\end{gather}
Now we separate the value of the field at the boundary $z=0$ by defining the bulk to boundary propagator $\calE_μ(ω,z)$ such that
\begin{gather}
    v_μ(ω,z) = \calE_μ(ω,z) v^0_μ(ω)
    \qquad\qquad
    \text{(no summation, $μ = 1,2,φ$),}
    \label{eq: bulk to boundary propagator}
\end{gather}
with $v^0_μ(ω) = v_μ(ω,0)$. This implies the bulk to boundary condition $\calE_μ(ω,0) = 1$. Using the definition \eqref{eq: bulk to boundary propagator} in equation \eqref{eq: on shell action without vt}, the on shell action becomes
\begin{gather}
    \begin{multlined}[c][0.93\textwidth]
    I_\onshell
    = -\ddfrac{1}{2g_5^2}
       ∫ḏω\, √{-g}\, ē^{-ϕ(z)} g^{zz} \,
           \LCHA[3] \sum_{j=1,2} g^{jj} \calE_j(-ω,z) v_j^0(-ω) \calE_j'(ω,z) v_j^0(ω) \\[-0.1\baselineskip]
                    + \COL[3]{\ddfrac{(g^{tφ})²}{-g^{tt}} + g^{φφ}} \calE_φ(-ω,z) v_φ^0(-ω) \calE_φ'(ω,z) v_φ^0(ω) \RCHA[3]
           \at[4]{}_{z\,=\,0}^{z\,=\,z_h}.
    \end{multlined}
\end{gather}
Then, applying the Son-Starinets prescription, we determine the retarded Green's functions
\begin{align}
    G_{jj}^R(ω) &= -\ddfrac{ℓ R}{g_5²} ē^{-ϕ(0)} \lim_{z-->0} \ddfrac{1}{z} \calE_j'(ω,z)
    \qquad \text{(no summation, $j=1,2$)}
    \label{eq: G_jj}
\shortintertext{and}
    G_{φφ}^R(ω) &= -\ddfrac{R}{ℓ g_5²} ē^{-ϕ(0)} \lim_{z-->0} \ddfrac{1}{z} \calE_φ'(ω,z).
    \label{eq: G_φφ}
\end{align}
The other $G_{μν}^R$ vanish.

In vacuum ($T = 0$), the Green's function is just
\begin{gather}
    Π(p²) = \sum_{n=1}^{∞} \ddfrac{f_n^2}{-p² - m_n² + îε},
\end{gather}
where $m_n$ and $f_n$ are the mass and decay constant of the radial states of excitation level  $n$ of bottomonium.
The imaginary part of this Green's function and, hence, the spectral function at zero temperature is proportional to
\begin{gather}
    \sum_{n=1}^{∞} f_n δ(p² + m_n²),
\end{gather}
a set of delta peaks, each one located at the mass $m_n$ of a state. 

When the meson is inside a thermal medium, at a non-zero temperature, the change in the spectral function is a broadening of the peaks. These peaks acquire a finite height and a non zero width. This broadening effect rises with the temperature and with the excitation level $n$ and is interpreted as dissociation in the thermal medium. Variations of the spectral function of quarkonia in a thermal medium without rotation can be found on 
\cite{Braga:2017bml,Braga:2018zlu, Braga:2018hjt,Braga:2019yeh}.
The same holographic model considered here was used in these references.

\subsection{Numerical Results of Spectral Functions}

\begin{figure}[htb!]
    \centering
    \vspace{\baselineskip}
    \includegraphics[scale=.75]{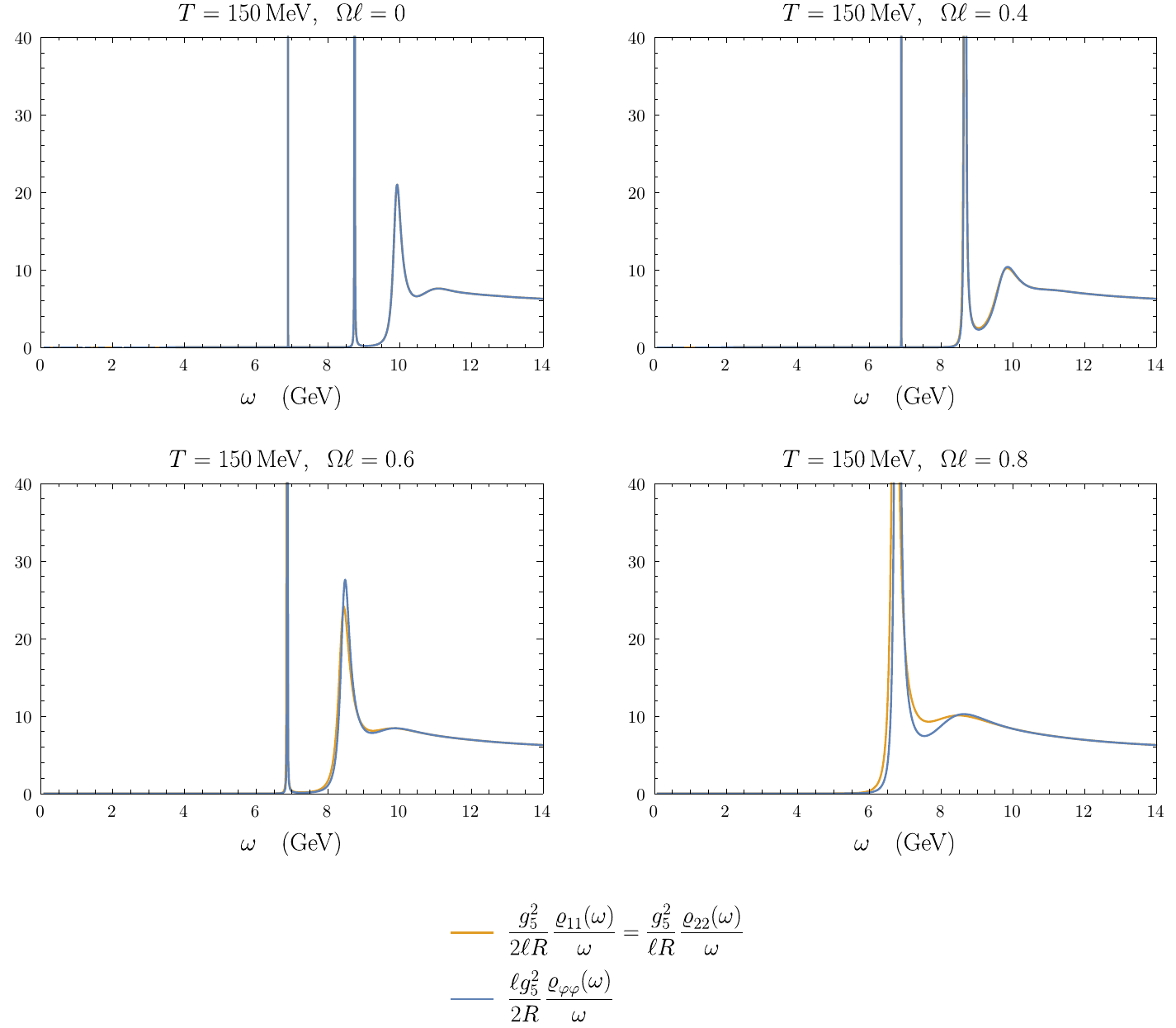}
    \caption{Spectral functions $ϱ_{11}^{}(ω) = ϱ_{22}^{}(ω)$ and $ϱ_{φφ}(ω)$ for different values of rotation speed $Ωℓ$ and temperature fixed at $T = \SI{150}{\mega\eV}$.}
    \label{fig: spectral function 150MeV}
\end{figure}

\begin{figure}[htb!]
    \centering
    \vspace{\baselineskip}
    \includegraphics[scale=.75]{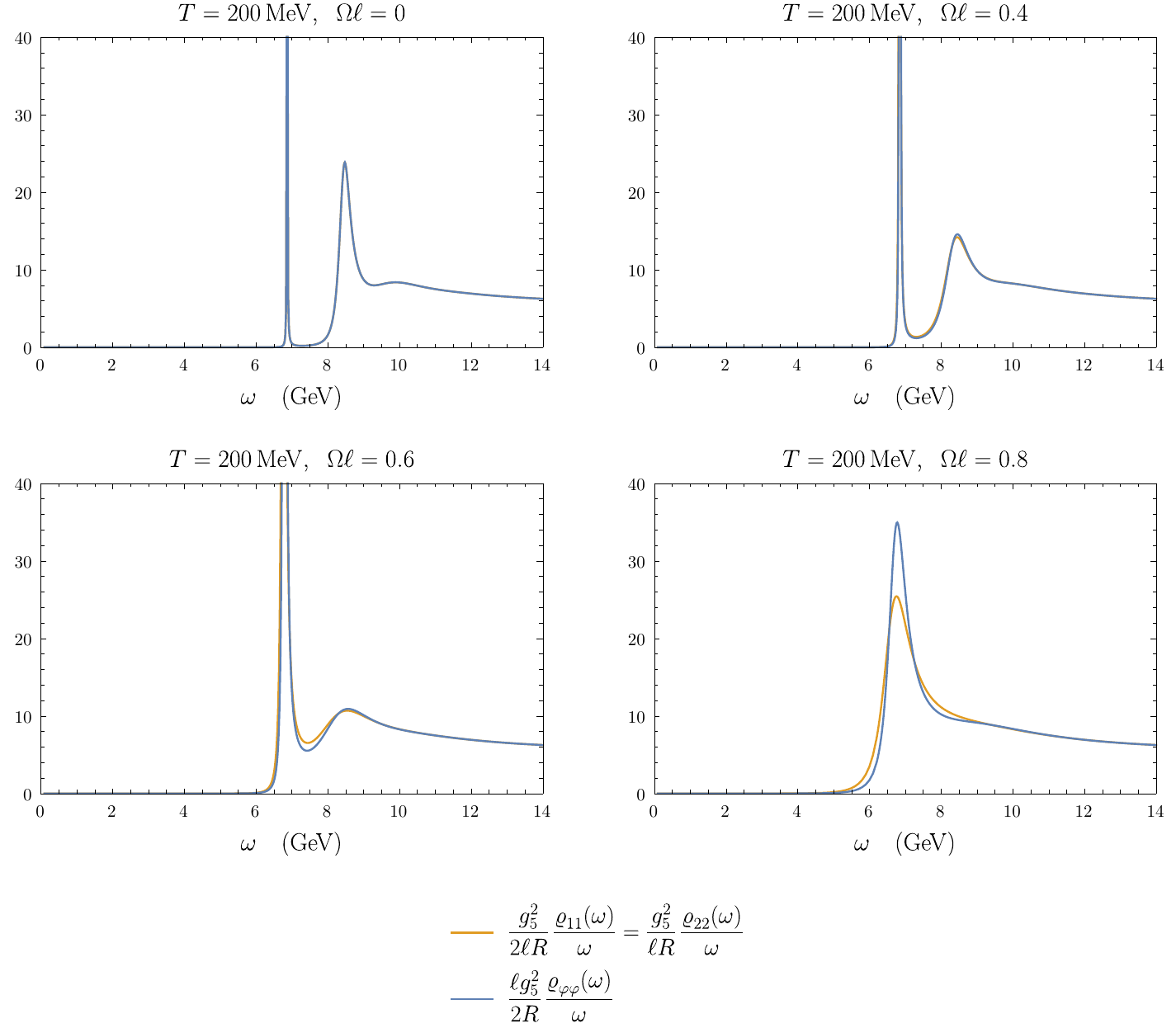}
    \caption{Spectral functions $ϱ_{11}^{}(ω) = ϱ_{22}^{}(ω)$ and $ϱ_{φφ^{}}(ω)$ for different values of rotation speed $Ωℓ$ and temperature fixed at $T = \SI{200}{\mega\eV}$.}
    \label{fig: spectral function 200MeV}
\end{figure}

\begin{figure}[htb!]
    \centering
    \vspace{\baselineskip}
    \includegraphics[scale=.75]{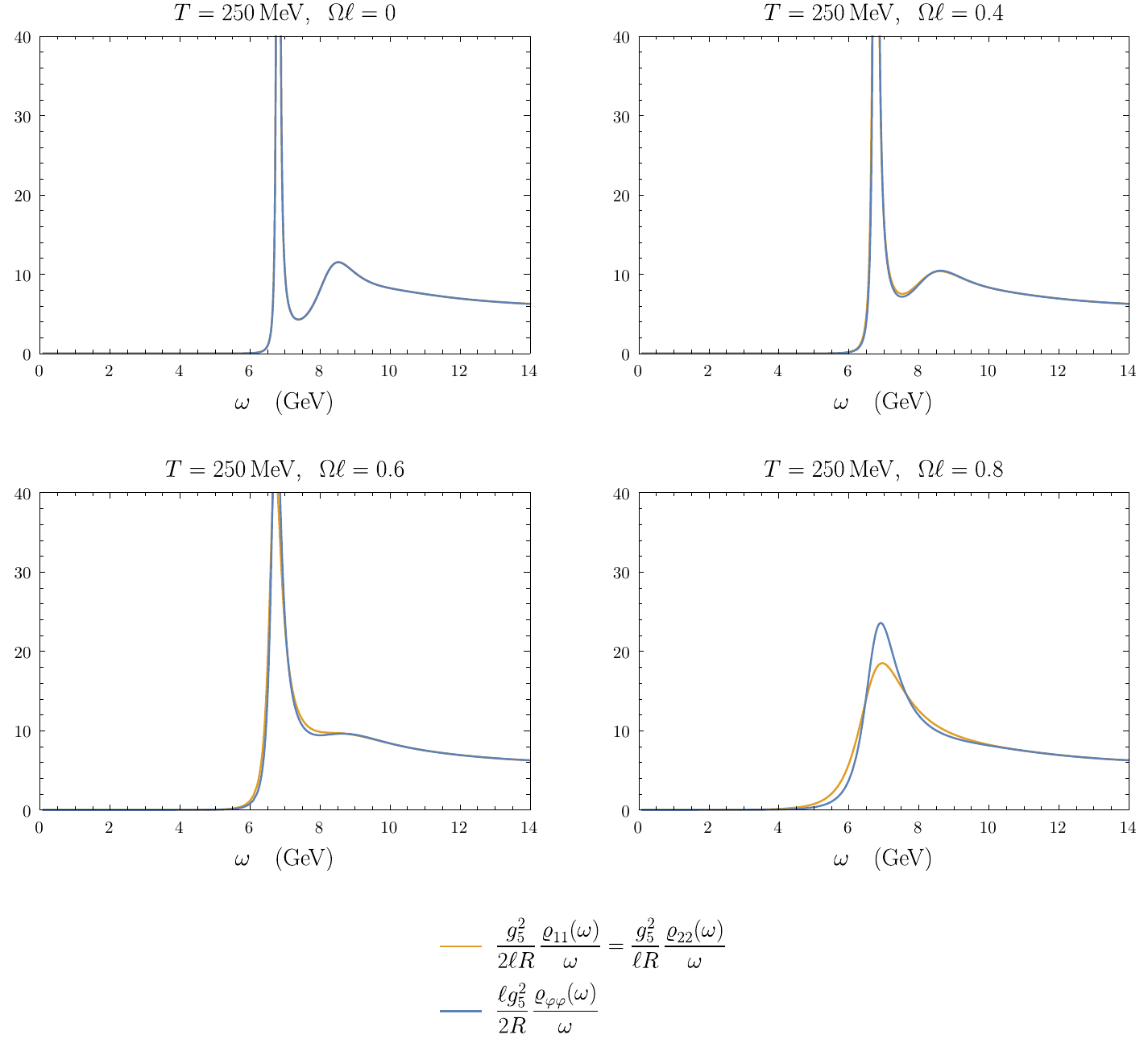}
    \caption{Spectral functions $ϱ_{11}^{}(ω) = ϱ_{22}^{}(ω)$ and $ϱ_{φφ}(ω)$ for different values of rotation speed $Ωℓ$ and temperature fixed at $T = \SI{250}{\mega\eV}$.}
    \label{fig: spectral function 250MeV}
\end{figure}

Figure \ref{fig: spectral function 150MeV} shows how bottomonium's spectral function changes with the rotation speed $Ω ℓ$ at temperature $T = \SI{150}{\mega\electronvolt}$. Figures \ref{fig: spectral function 200MeV} and \ref{fig: spectral function 250MeV} do the same for temperatures fixed at $T = \SI{200}{\mega\electronvolt}$ and $T = \SI{250}{\mega\electronvolt}$, respectively. In these charts we multiplied the spectral functions by the inverse of the constants that appear in eqs. \eqref{eq: G_jj} and \eqref{eq: G_φφ} in order to represent functions with the same dimension, that can be compared. From these figures one can see that rotation increases the dissociation effect and also that fields with polarization $v_1$ and $v_2$ dissociate slightly faster than the ones with polarization $v_φ$.


\vspace{.75\baselineskip}
\section{Quasinormal Modes}
\label{sec: QNMs}

In vacuum, the equations of motion simplify to
\begin{gather}
    ω² v + \PAR[3]{\!\!- \ddfrac{1}{z} - ϕ' } v' + v'' = 0.
\end{gather}
In this case, there is no black hole and, therefore, no infalling wave condition. We determine the normal modes by solving these equations with the exigence of the field to satisfy the normalization condition
\begin{gather}
    ∫_0^∞ \dfrac{R}{z} ē^{-ϕ(z)}\, |v(z)|²\, ḏz = 1.
\end{gather}
It is possible to translate this normalization condition into the Dirichlet condition
\begin{gather}
    v(ω, z = 0) = 0.
    \label{def: QNM}
\end{gather}
This equation is solvable only for a discrete set of real values $ω_n$. These values are the masses of the quarkonium states in vacuum. They are shown in the third column of table \ref{tab: bottomonium masses and decay constants}.

At finite temperature, instead of normal modes, we have the quasinormal modes. They are the solutions of the equations of motion \eqref{eq: eq motion v_1,2} and \eqref{eq: eq motion v_φ}, that satisfy
\begin{enumerate}
\setstretch{.75}
\item the infalling wave condition at the horizon,
\item the Dirichlet condition \eqref{def: QNM}.
\vspace{-.75\baselineskip}
\end{enumerate}
The values $ω_n$ that satisfy both of this conditions are called quasinormal frequencies, the fields $v_μ(ω_n, z)$ are called quasinormal modes and represent the meson quasistates.

As the value at $T = 0$ of the normal frequency $ω_n$ is interpreted as the mass of the particle in its state $n$, the real part of the quasinormal frequency $\Re(ω_n)$, at finite temperature, is interpreted as the thermal mass of the quasiparticle. The imaginary part $\Im(ω_n)$ is related to its degree of dissociation. The larger the absolute value of the imaginary part, the stronger the dissociation. It is interesting to note that the real and imaginary parts of a $n$-th quasinormal mode are related to the position and width of the $n$-th peak in the spectral function. Therefore, we can interpret a growth in the imaginary part of the quasinormal frequency as an increase in the dissociation effect. Indeed, at $T = 0$ the width of the spectral function peaks is zero, as is the imaginary part of the frequency $ω_n$. Also, the limit for $T -> 0$ of $\Re(ω_n)$ is the mass $m_n$. This discussion for the non-rotating plasma with temperature only is already present in the literature. One can find an application of the tangent model for this case on Refs.\cite{Braga:2017bml,Braga:2018zlu, Braga:2018hjt,Braga:2019yeh}.

\begin{figure}[htb!]
    \centering
    \vspace{\baselineskip}
    \includegraphics[scale=.75]{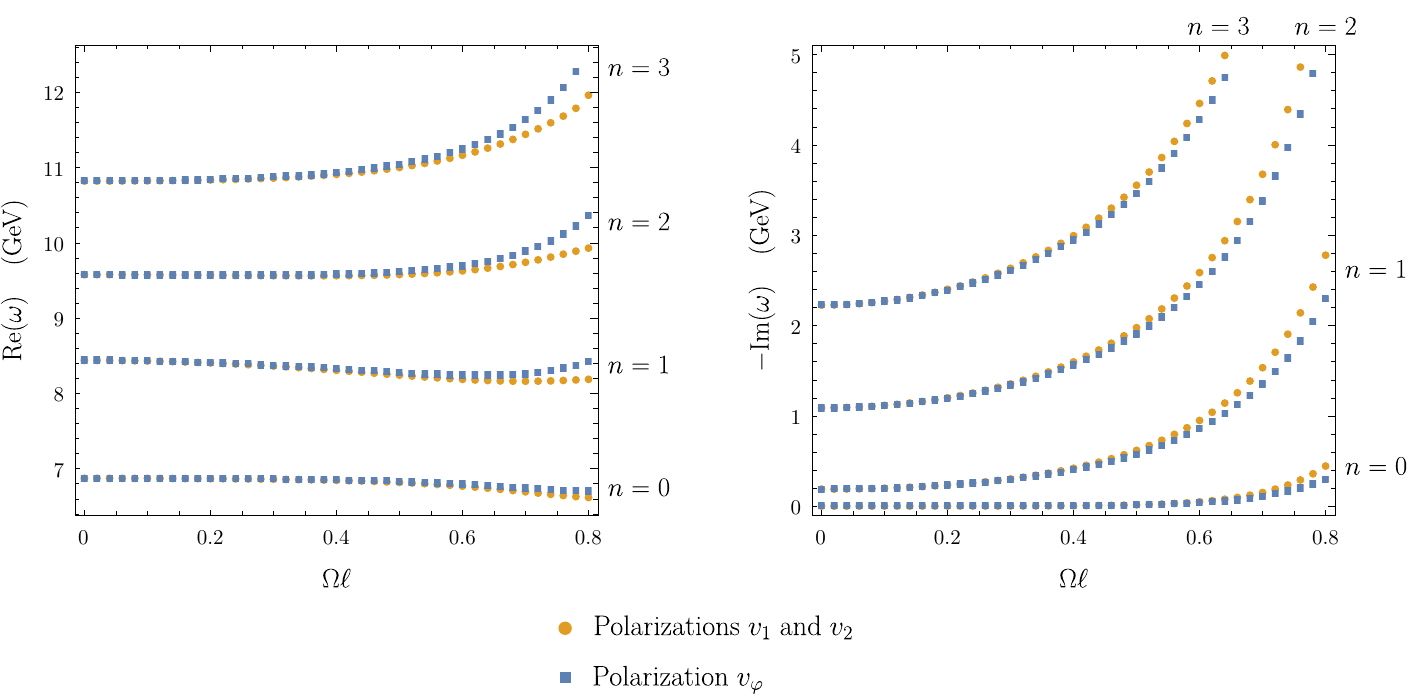}
    \caption{Quasinormal frequencies of the different excitation levels as functions of the rotation speed $Ωℓ$ and temperature fixed at $T = \SI{200}{\mega\eV}$.}
    \label{fig: QNMs}
\end{figure}

The results of quasinormal frequencies for polarizations $x^j$ and $φ$ as function of the rotation speed $Ωℓ$ and for temperature fixed at $T = \SI{200}{\mega\eV}$ are shown in figure \ref{fig: QNMs}. From this figure, one sees that the dissociation degree, measured by $-\Im(ω)$,  rises with the rotation speed $Ωℓ$.

\vspace{.75\baselineskip}
\section{Conclusions}
\label{sec: Conclusions}
We analysed in this work how does rotation of a quark gluon plasma affect the dissociation of heavy vector mesons that are inside the medium. The motivation for such a study is the fact that non central heavy ion collisions lead to the formation of a QGP with high angular momentum. So, a description of 
 quarkonium inside the plasma should take rotation into account. We considered, as an initial study, the case of a cylindrical shell of plasma in rotation about the symmetry axis. The real case of the QGP should involve a volume rather than a cylindrical surface and also possible interaction between different layers of the plasma, that would have different rotational speeds. However this simple case considered here already provides important non trivial information. It is clear from the results obtained here that
rotation enhances the dissociation process for heavy vector mesons inside a plasma. It was also found that these effect, caused by rotation, is more intense for heavy vector mesons that have polarization perpendicular to the rotation axis.

\noindent {\bf Acknowledgments:} N.R.F.B. is partially supported by CNPq --- Conselho Nacional de Desenvolvimento Científico e Tecnologico grant 307641/2015-5 and by FAPERJ --- Fundação Carlos Chagas Filho de Amparo à Pesquisa do Estado do Rio de Janeiro. The authors received also support from Coordenação de Aperfeiçoamento de Pessoal de Nível Superior --- Brasil (CAPES), Finance Code 001.

\setstretch{1.2}
\bibliography{bibliography}

\end{document}